\documentclass[fleqn,12pt,twoside]{article}
\usepackage[headings]{espcrc1}

\usepackage{xspace}
\usepackage[tight]{subfigure}

\readRCS
$Id: QM2005_Proceedings.tex,v 1.7 2005/09/30 02:15:50 winter Exp $
\usepackage{graphicx}

\usepackage[figuresright]{rotating}

\newcommand{\AmS}{{\protect\the\textfont2
  A\kern-.1667em\lower.5ex\hbox{M}\kern-.125emS}}

\newcommand{\pT}{\ensuremath{p_T}\xspace}
\newcommand{\vTwo}{\ensuremath{v_2}\xspace}
\newcommand{\vTwoRaw}{\ensuremath{v_2^{raw}}\xspace}
\newcommand{\vTwoCorr}{\ensuremath{v_2^{corr}}\xspace}
\newcommand{\piz}{\ensuremath{\pi^0}\xspace}
\newcommand{\RAA}{\ensuremath{R_{AA}}\xspace}
\newcommand{\photon}{\ensuremath{\gamma}\xspace}
\newcommand{\meanNpart}{\ensuremath{\langle N_{part}\rangle}\xspace}
\newcommand{\dphi}{\ensuremath{\Delta\phi}\xspace}
\newcommand{\leff}{\ensuremath{L_{{\it eff}}}\xspace}

\makeatletter
\newcommand*{\SubLabels}[1]{%
  \label{fig:#1}%
  \begingroup
    \protected@edef\@currentlabel{%
      \csname thesub\@captype\endcsname
    }%
    \label{subfig:#1}%
  \endgroup
}
\makeatother

\title{PHENIX Measurement of Particle Yields at High $p_T$ with Respect to
Reaction Plane in Au+Au Collisions at $\sqrt{s_{NN}} = 200$~GeV}

\author{David Winter\address[MCSD]{Department of Physics / Nevis Laboratories, 
        Columbia Unviersity, \\ 
        538 West 120th St., New York, NY 10027, USA}
	(for the PHENIX\thanks{For the full list of PHENIX
        authors and acknowledgements, see Appendix 'Collaborations' of this volume} Collaboration)}
       
\runtitle{High $p_T$ Yields with Respect to Reaction Plane in 200 GeV
Au+Au Collisions} \runauthor{D. Winter}

\begin{document}

\maketitle

\begin{abstract}
The PHENIX Run 4 Au+Au dataset provides a powerful opportunity for
exploring the angular anisotropy of identified particle yields at high
\pT.  Complementing traditional \vTwo measurements, we present \piz
yields as a function of angle with reaction plane, up to $p_T$ 10
GeV/c. The centrality dependence of the angular anisotropy allows us
to probe the density and path-length dependence of the energy loss of
hard-scattered partons.  We will discuss various mechanisms for
particle production in this high $p_T$ region.
\end{abstract}

\section{Introduction}

One of the most intriguing puzzles in RHIC physics is the origin of
the azimuthal anisotropy of particle yields at high $p_T$
($>5$~GeV/c)~\cite{ref:highptv2,ref:gvw}. Parton energy loss, which
can reproduce the flat, suppressed \RAA in single particle
spectra~\cite{ref:tadaaki_pizRaa}, fails to reproduce the required
\vTwo at intermediate \pT~\cite{ref:shuryak,ref:dfj}.  The complicated
interplay between soft and hard physics (for example, recombination,
flow, and energy loss) at intermediate \pT makes it critical to
measure \vTwo at high \pT, where we expect the only contribution to be
from energy loss.  We report here on the measurement of the \piz
azimuthal asymmetry, and discuss what it tells us about the source(s)
of high \pT \vTwo.  First we describe the method by which we measure
\RAA as a function of angle with respect the reaction plane and \vTwo
for \piz{}s in PHENIX.  We then discuss two model calculations that
help to shed light on the mechanisms giving rise to the high \pT
\vTwo.



\section{Measuring \piz yields and \vTwo in PHENIX}

During Run 4, PHENIX recorded 1.5 billion minimum bias Au+Au events at
$\sqrt{s} = 200$~GeV/c; the data presented here is taken from 1
billion of those events.  For measuring photons and \piz{}s, we use
the Electomagnetic Calorimeter (EmCal)~\cite{ref:emcal}.  The angle of
the reaction plane is measured event-by-event using the Beam-Beam
Counters.  Pairs of clusters that pass \photon identification cuts are
binned in angle with respect to the reaction plane ($\Delta\phi = \phi
- \Psi_{RP}$).  A similarly binned mixed event background is then
subtracted. The counts in the remaining peak are integrated in a $\pm
2\sigma$ window, determined by a gaussian fit.  Six bins in
$\Delta\phi$ are used from $0-\pi/2$.

To measure \vTwo, we fit the $\Delta\phi$ distribution as $1 + 2\vTwoRaw
\cos(2\Delta\phi)$.  The resulting \vTwo parameter needs to be
corrected for the reaction plane resolution, hence the
designation \vTwoRaw.  The resolution $\sigma$ is determined for each
centrality bin, and leads to the corrected value $\vTwoCorr =
\vTwoRaw/\sigma$.  The yields as a function of $\Delta\phi$ can then
be corrected with a factor
\begin{equation}
f = \frac{1+2\vTwoCorr\cos2\Delta\phi}{1+2\vTwoRaw\cos2\Delta\phi}.
\end{equation}

\section{Results and Discussion}

\begin{figure}[t]
\begin{center}
  \subfigure[\SubLabels{raa_npart}\piz \RAA
    vs. $N_{part}$]{\includegraphics[width=0.48\textwidth]{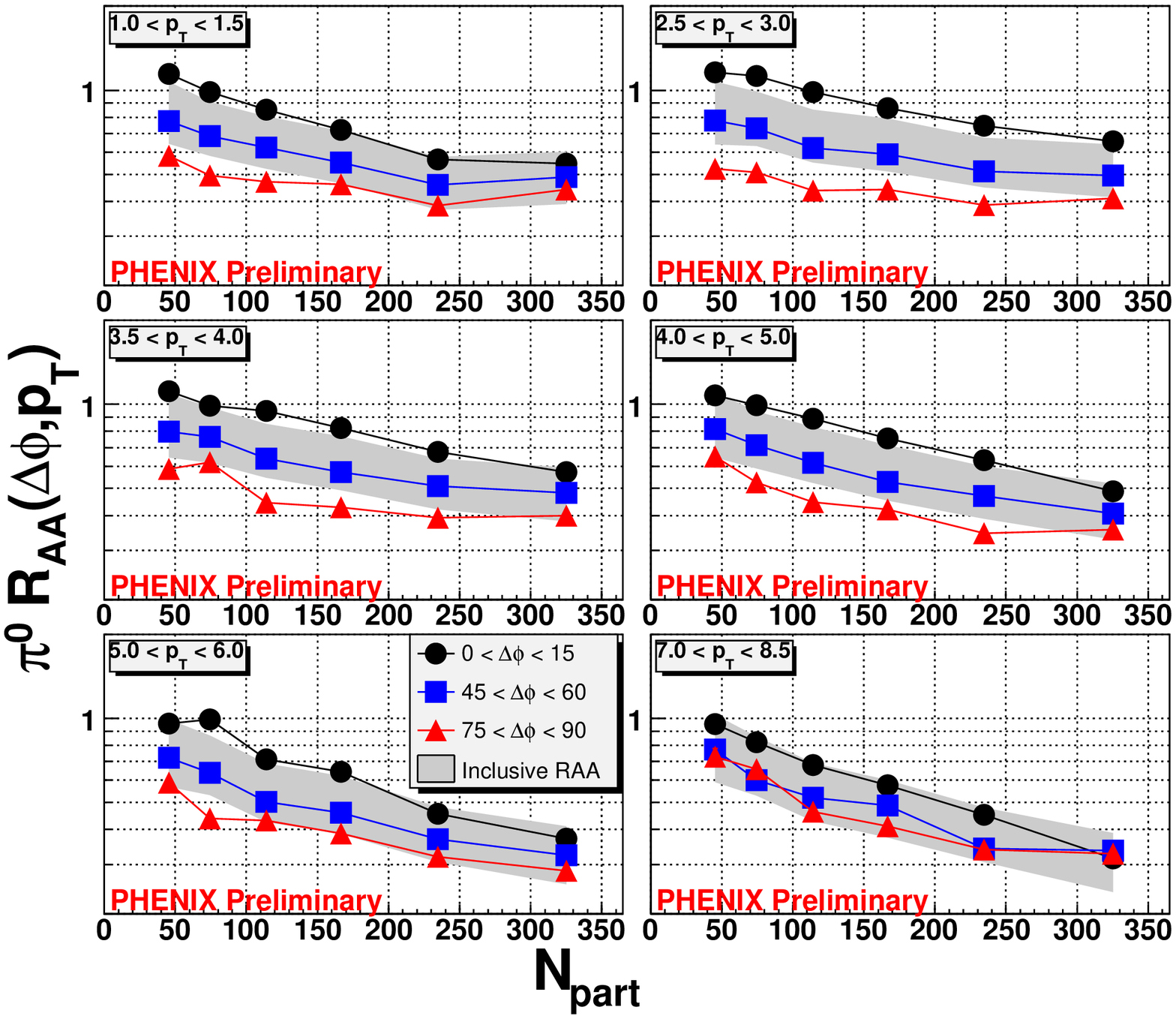}
  }
  \subfigure[\SubLabels{piz_v2}\piz \vTwo
    vs. \pT]{\includegraphics[width=0.48\textwidth]{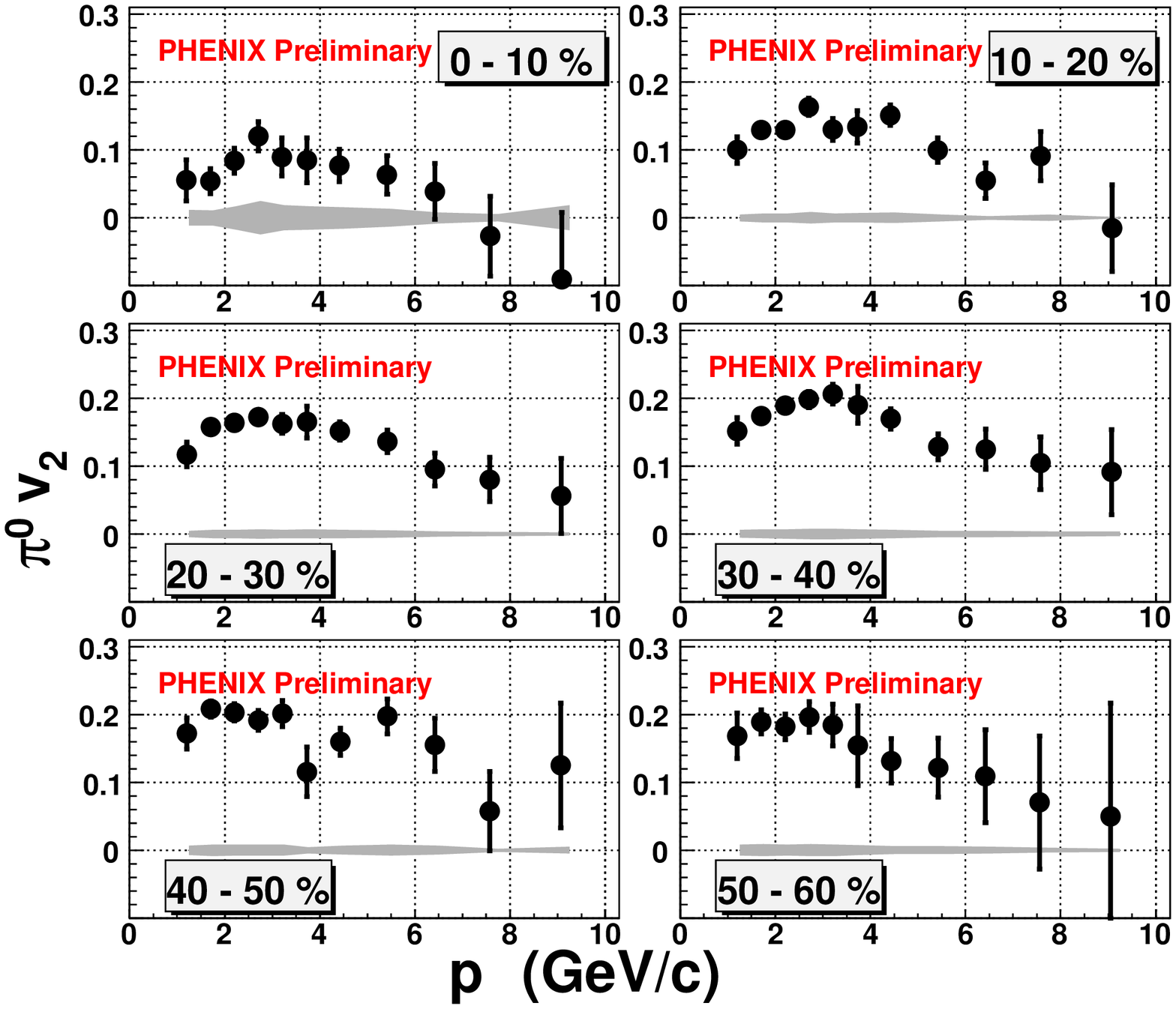}
  }
\end{center}
\vspace{-20pt}
\caption{\piz \vTwo and \RAA:  \ref{subfig:raa_npart} shows $\RAA(\pT)$
  as a function of \meanNpart for each \dphi bin; the panels are for
  different \pT bins.  \ref{subfig:piz_v2} the \piz $\vTwo(\pT)$, with
  each panel corresponding to a centrality bin.  The grey bands indicate the
  systematic error due to the reaction plane resolution correction.}
\end{figure}

To obtain $\RAA(\Delta\phi)$, we exploit the fact that the ratio of
of the yield at a given $\Delta\phi$ to the inclusive yield is
equivalent to the ratio of the angle-dependent \RAA to the inclusive
\RAA.  Thus mulitplying these relative yields by an inclusive measured
$\RAA$, we have:
\begin{equation}
\RAA(\Delta\phi) = {\tt Yield}(\Delta\phi) / {\tt Yield} \times \RAA
\end{equation}

The $\RAA(\Delta\phi,\pT)$ as a function of \meanNpart is shown in
Figure~\ref{fig:raa_npart}.  It is clear that there is non-trivial
substructure to the angular dependence of the \RAA, and that it varies
with centralities.  This feature is emphasized by plotting the data on
a semi-log scale, showing that the \RAA behaves differently in
different \dphi bins.

The resulting \piz \vTwo is shown in Figure~\ref{fig:piz_v2}.  For the
first time we observe \vTwo up to 10 GeV/c, and we see a clear
decrease at high \pT, then a leveling off to a finite value.

\begin{figure}[t]
\begin{center}
\includegraphics*[width=\textwidth]{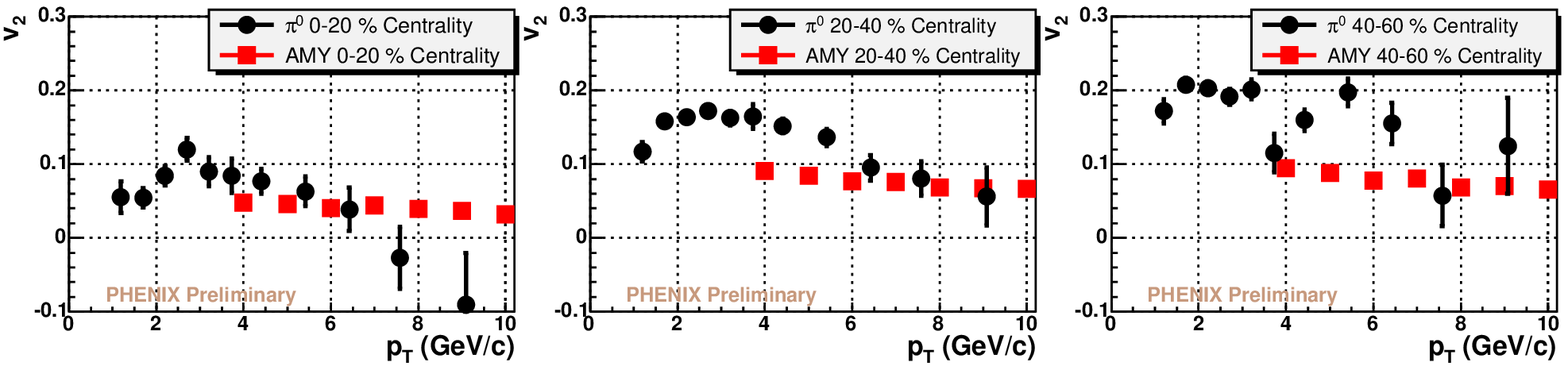}
\includegraphics*[height=4cm]{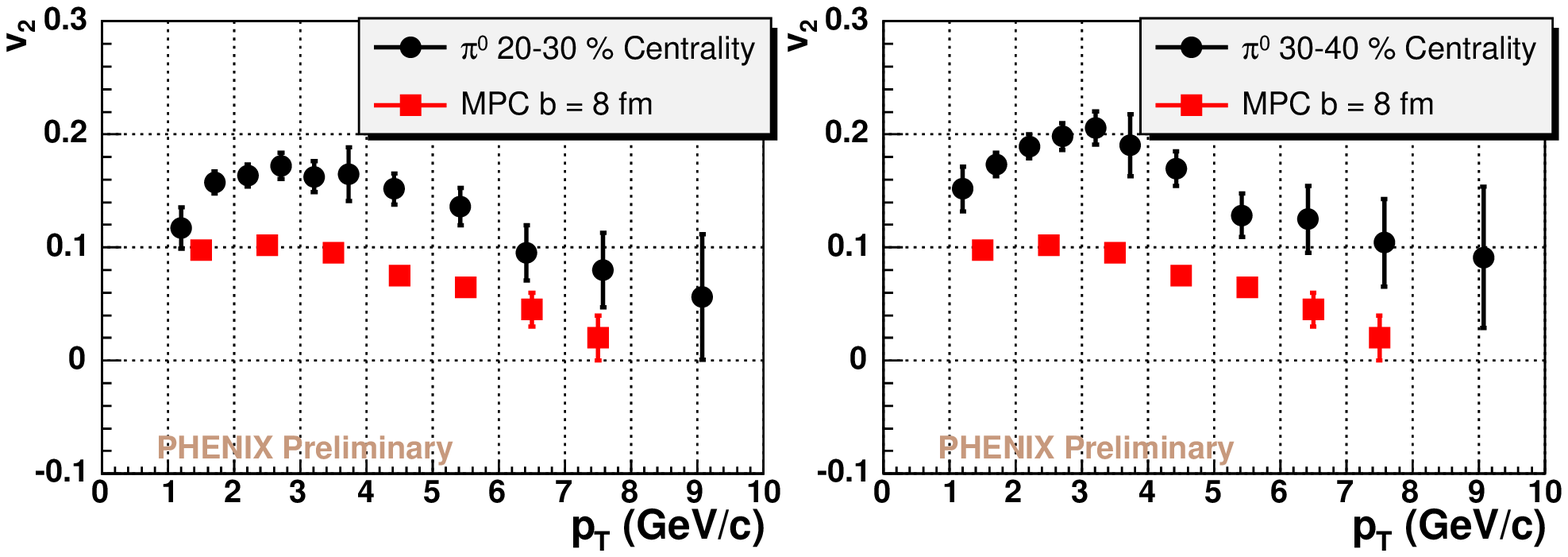}
\end{center}
\vspace{-20pt}
\caption{Comparison of \piz \vTwo with models. The top three panels show the
  AMY calculation with data for three centralities.  The bottom two panels
  compare two centralities with the $b=8$~fm calculation of the MPC model.}
\label{fig:modelCompare}
\end{figure}

To gain insight into the \vTwo mechanisms at work at high \pT, we turn
to models.  We compare the \piz \vTwo to two models, an
Arnold-Moore-Yaffe (AMY) calcuation~\cite{ref:amy} done by Turbide et
al. and the Molnar Parton Cascade (MPC) model~\cite{ref:mpc}.
Figure~\ref{fig:modelCompare} shows calculations from these models,
plotted alongside data for similar centralities.  The AMY calculation
contains energy loss mechanisms only, and we see that the data appear
to decrease to a value at high \pT that is consistent with this model;
the agreement is most striking in the 20-30\% bin.

The MPC model has a number of mechanisms in it, including corona
effects, energy loss, and the ability to boost lower \pT partons to
higher \pT (a unique feature).  The calculation shown in
Figure~\ref{fig:modelCompare} does a better job of reproducing the
overall shape of the \vTwo, though it is systematically low.  It is
important to note that this calculation is done for one set of
parameters, so it should be very interesting to see if the MPC can
better reproduce the data for a different set of parameters.

The prevailing mechanism for high \pT \vTwo is parton energy loss in
the dense medium.  If this is true, the \RAA should be sensitive only
to the geometry of the collision.  To test this behavior, we seek to
reparamaterize the two handles we have on geometry (centrality, or
collision overlap, and angle of emission) into a single parameter, an
effective path length which we will refer to as $\leff$.  Details of
the calculation are described in~\cite{ref:leff}.  In essence, it is
propotional to the parton-density weighted average of the length from
origin to edge of an ellipse.  We also perform a Glauber Monte Carlo
sampling of starting points to account for fluctuations in the
location of the hard-scattering origin of the particles' paths.

\begin{figure}[t]
\begin{center}
\includegraphics*[width=\textwidth]{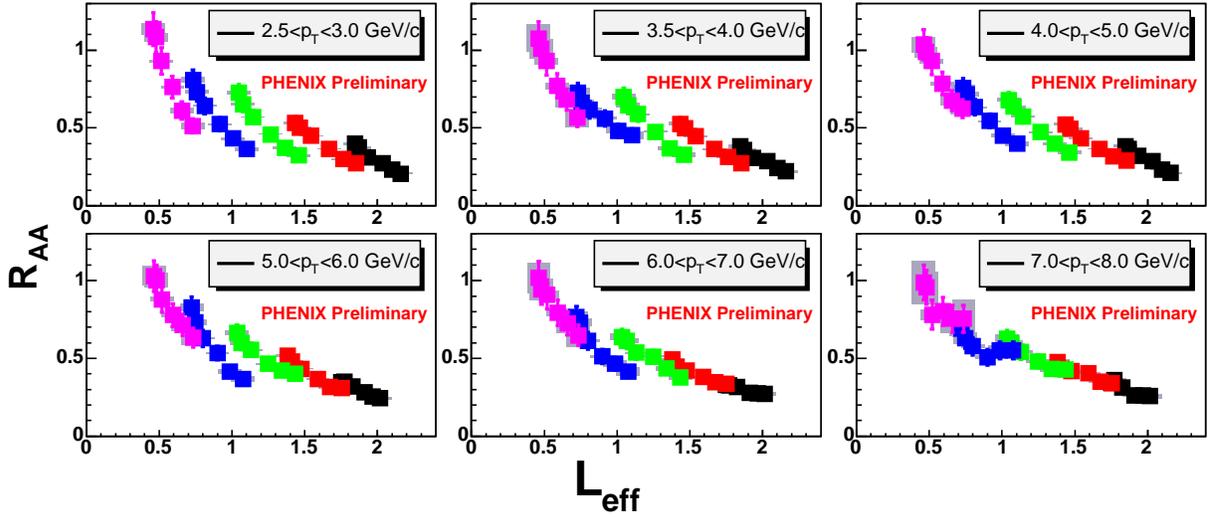}
\end{center}
\vspace{-20pt}
\caption{$R_{AA}(\dphi,\pT)$ vs. $\leff$.  The panels correspond to
  different \pT ranges.  The colors represent the centrality bins:
  magenta is most peripheral while black is most central.}
\label{fig:leff}
\end{figure}

The result of plotting \RAA for all centralities and angles
vs. $\leff$ is shown in Figure~\ref{fig:leff}.  If the observed
\RAA arose from only geometric effects, we would expect the data to
exhibit a universal dependence on \leff.  For low \pT, this is clearly
not the case; something more than just energy loss is taking place
there.  However, when the \pT reaches 7 GeV/c and above, the \RAA data
do indeed appear to have a dependence on a single \leff curve.

\section{Summary}

We have presented the first measurement of high \pT \vTwo for \piz and
charged hadrons.  It is now clear that the \vTwo at $\pT > 7$~GeV/c
decreases to a small but non-trivial value.  Comparison with models
suggest that the dominant mechanism at work at high \pT is energy
loss.  In addition, we have also presented the first measurement of
\piz \RAA as a function of angle with respect to the reaction plane.
The $\RAA(\dphi,\pT)$ exhibits interesting angular substructure.
Furthermore, when the \RAA data are plotted as a function of an
effective path length through the medium, they exhibit a universal
behavior at $\pT > 7$~GeV/c.



\end{document}